# Leakage-free electrolytes with different conductivity for non-volatile memory device utilizing insulator/metal ferromagnet transition of SrCoO$_x$

Takayoshi Katase[1,2,3,a], Yuki Suzuki[4], and Hiromichi Ohta[1,a]

[1] Research Institute for Electronic Science, Hokkaido University, N20W10, Kita, Sapporo, 001-0020, Japan

[2] Laboratory for Materials and Structures, Institute of Innovative Research, Tokyo Institute of Technology, 4259 Nagatsuta, Midori, Yokohama, 226−8503, Japan

[3] PRESTO, Japan Science and Technology Agency, 7 Gobancho, Chiyoda, Tokyo, 102-0076, Japan

[4] School of Information Science and Technology, Hokkaido University, N14W19, Kita, Sapporo, 060−0814, Japan

[a] Correspondence and requests for materials should be addressed to: T. Katase (katase@mces.titech.ac.jp) and H. Ohta (hiromichi.ohta@es.hokudai.ac.jp)

**Abstract**

The electrochemical switching of SrCoO$_x$-based non-volatile memory with thin-film-transistor structure was examined by using liquid-leakage-free electrolytes with different conductivity ($\sigma$) as the gate insulator. We first examined leakage-free water, which is incorporated in the amorphous (a-) 12CaO·7Al$_2$O$_3$ film with nanoporous structure (CAN), but the electrochemical oxidation/reduction of SrCoO$_x$ layer required the application of high gate voltage ($V_g$) up to 20 V for a very long retention-time ($t$) ~40 minutes, primarily due to the low $\sigma$ (2.0 × 10$^{-8}$ S cm$^{-1}$ at RT) of leakage-free water. We then controlled the $\sigma$ of leakage-free electrolyte, infiltrated in the a-Na$_x$TaO$_3$ film with nanopillar array structure, from 8.0 × 10$^{-8}$ S cm$^{-1}$ to 2.5× 10$^{-6}$ S cm$^{-1}$ at RT by changing the $x$ = 0.01–1.0. As the result, the $t$, required for the metallization of SrCoO$_x$ layer under small $V_g$ = –3 V, becomes two orders of magnitude shorter with increase of the $\sigma$ of the a-Na$_x$TaO$_3$ leakage-free electrolyte. These results indicate that the ion migration in the leakage-free electrolyte is the rate-determining step for the electrochemical switching, compared to the other electrochemical process, and the high $\sigma$ of the leakage-free electrolyte is the key factor for the development of the non-volatile SrCoO$_x$-based electro-magnetic phase switching device.

# I. Introduction

Transition metal oxides (TMOs) have been expected as an active material for the novel functional devices, because they exhibit the rich variety of physical properties, such as metal-insulator transition, magnetic transition, electrochromism, and superconductivity.[1-5] The variation of the functional properties in TMOs originates from the flexibility of valence state of TM ions.[6] By changing the number of $d$ electrons in TM ions, the TMOs show the dramatic changes in the optical, electronic, and magnetic properties through phase transitions.[7]

Strontium cobaltite ($SrCoO_x$) shows the distinct electro-magnetic properties depending on the valence state of Co ion.[8,9] The oxygen off-stoichiometry ($x$) in $SrCoO_x$ can be varied in the range of 2.5–3.0, where the oxygen-vacant $SrCoO_{2.5}$ phase has a brownmillerite (BM-) type structure and the fully-oxidized $SrCoO_3$ phase has a simple perovskite (P-) type structure. The BM-$SrCoO_{2.5}$ with $Co^{3+}$ ($3d^6$) is an antiferromagnetic (AFM) insulator with Néel temperature of 570 K,[8] whereas the P-$SrCoO_3$ with $Co^{4+}$ ($3d^5$) is a ferromagnetic (FM) metal with Curie temperature of 275 K.[10] Their highly contrastive ground states between AFM insulator and FM metal should be promising for the novel electro-magnetic phase switching devices.

In order to modulate their electron-magnetic properties, there are two approaches for the control of $x$ in $SrCoO_x$. High-temperature heat treatment under oxidative/reductive conditions has been a conventional way to control the $x$ of TMOs.[11-13] For $SrCoO_x$ thin films, the two phases could be altered reversibly at least 200–300 °C,[14] but the heating process for the phase switching is unsuited for the device operating at room temperature (RT). Meanwhile, the redox reaction using liquid electrolyte is another classical method for the control of $x$ at RT;[15] the reversible redox reaction of $SrCoO_x$ has been accomplished by the electrolysis in alkaline aqueous solutions.[16-19] However, it cannot be used for practical purposes without sealing; the electro-magnetic phase switching should be operatable in a solid-state device at RT.

In order to overcome this issue, we have proposed liquid-leakage-free electrolytes incorporated in nanoporous insulators. In 2010, we have developed an amorphous (a-) $12CaO·7Al_2O_3$ film with nanoporous structure (Calcium Aluminate with Nanopore, CAN) as a leakage-free water.[20] The CAN is a glassy solid film containing nanopores with average diameters of 10–20 nm. Water vapor in air is automatically absorbed into the nanopores due to capillary action and the percolation conduction of water can be observed in the interconnected nanopores. We further developed an a-$NaTaO_3$ film with nanopillar array structure, which works as a leakage-free electrolyte containing an alkali hydroxide as a salt.[21] Also for this case, the water vapor in air is absorbed into the interspace of nanopillar arrays and then, a small amount of $Na^+$ ions from the a-$NaTaO_3$ nanopillars is dissolved into the adsorbed water, leading to the NaOH aqueous solution infiltrated in the solid a-$NaTaO_3$ film. Thus, no leakage of the electrolyte occurs due to the large surface tension of the electrolyte/nanopores; i.e. the formation of liquid-leakage-free electrolyte.

Recently, we demonstrated an electrochemically switchable electro-magnetic phase switching device of $SrCoO_x$ by using a three terminal TFT structure with the leakage-free electrolyte.[21-23] The TFT structure is composed of $SrCoO_x$ film as an active channel layer and a-$NaTaO_3$ film with nanopillar array structure as a gate insulator. By applying the gate voltage, the electrochemical oxidation/reduction occurs and the reversible switching between AFM insulator and FM metal of the $SrCoO_x$ layer was realized under a small DC voltage (±3 V) within a very short time (2–3 sec.) at RT in air.[21] The electrochemical reaction of SrCoO$_x$ layer obeys the Faraday's laws of electrolysis and the device operation can be controlled by current density flowing in the device. Therefore it is considered that the conductivity of the leakage-free electrolyte is the key factor for the electrochemical switching of SrCoO$_x$-based TFT, but the conductivity control of the leakage-free electrolyte and their effects on the device operation has not been examined yet; the systematic characterization should lead to the further understanding for the electrochemical switching of SrCoO$_x$-based TFT.

In this paper, we examined the electrochemical switching of SrCoO$_x$ TFT by using liquid-leakage-free electrolytes of CAN film and a-Na$_x$TaO$_3$ films with different conductivity. We fabricated the a-Na$_x$TaO$_3$ films with different $x$ = 0.01–1.0, to demonstrate the conductivity control of the leakage-free electrolyte. As the result, the conductivity of a-Na$_x$TaO$_3$ films can be largely controlled from $8.0 \times 10^{-8}$ S cm$^{-1}$ ($x$ = 0.01) to $2.5 \times 10^{-6}$ S cm$^{-1}$ ($x$ = 1) at RT, compared to the $2.0 \times 10^{-8}$ S cm$^{-1}$ of the CAN film. We examined the effect of the different conductivity of the leakage-free electrolytes on the device operation.

## II. Experimental

### A. Fabrication of leakage-free electrolytes

The leakage-free electrolytes of CAN film and Na$_x$TaO$_3$ films were fabricated by pulsed laser deposition (PLD) at RT. A KrF excimer laser (wavelength of 248 nm, repetition rate of 10 Hz) was used to ablate the targets of 12CaO·7Al$_2$O$_3$ and Na$_x$TaO$_3$ ceramics. In order to control the Na concentration ($x$) in the Na$_x$TaO$_3$ films, the chemical composition of the targets of Na$_x$TaO$_3$, which are mixture of NaTaO$_3$ and Ta$_2$O$_5$ phases, was changed in the range of $x$ = 0.01–1.0.

## B. Fabrication of SrCoO$_x$-TFT with leakage-free electrolyte

**Figure 1** schematically illustrates the TFT structure with 800-μm-long and 400-μm-wide channel, which is fabricated using stencil masks onto an atomically-flat stepped-and-terraced surface of SrCoO$_{2.5}$ epitaxial film;[21] the SrCoO$_{2.5}$ layer was deposited on (001) SrTiO$_3$ single crystal (substrate size: 10×10×0.5 mm$^3$) at 720 °C under oxygen pressure ($P_{O2}$) of 10 Pa by PLD. Cr/Au bilayer film, which served as source and drain electrodes for the ohmic contant to SrCoO$_x$ layer, was deposited by electron beam (EB) evaporation at RT. The liquid-leakage-free electrolytes of CAN film and Na$_x$TaO$_3$ films were deposited on the SrCoO$_{2.5}$ layer. Finally, a-WO$_3$ counter layer and Ti gate electrode were respectively deposited by PLD and EB evaporation on the leakage-free electrolyte. By applying the negative gate voltage ($V_g$), OH$^−$ ion is attracted to the surface of SrCoO$_x$ layer and the electrochemical oxidation occurs. Meanwhile, by applying the positive $V_g$, the reverse reaction occurs for the reduction of SrCoO$_x$ layer. The a-WO$_3$ counter layer works as the H$^+$ absorber,[24] which maintains a better electrochemical balance in the device.

## C. Characterization

The film density of the leakage-free electrolytes was characterized by X-ray reflectivity measurement, and the film porosity ($\Phi$) was calculated from the ratio of the densities of porous and dense films. Cross-sectional microstructure of thin films was examined by high-resolution transmission electron microscopy (TEM) and scanning TEM (JEM-ARM200F, 200 kV, JEOL Ltd.). The cross-sectional samples for TEM observations were prepared by focused-ion-beam (FIB) micro-sampling technique, in which the multilayer structure region of the TFTs was cutout and thinned by FIB (FB-2000A, HITACHI).

The conductivity ($\sigma$) of the leakage-free electrolytes was measured by the AC impedance method using an impedance analyzer (YHP4192A, Yokogawa-Hewlett-Packard) and potentio/galvanostat (VersaSTAT 4, Princeton Applied Research ltd.), where the films were sandwiched by Au electrodes and the $\sigma$ was measured perpendicular to the thin-film plane. The $\sigma$ was estimated from the intercept on the real axis of the semicircle in the Cole-Cole plots.

The SrCoO$_x$-TFT characteristics were examined by measureing the sheet resistance ($R_s$) and thermopower ($S$), after applying $V_g$ and subsequently switching the $V_g$ off, at RT in air. The $V_g$ is applied between the gate-source electrodes of the device. The gate current ($I_g$) during the $V_g$ application was measured by a source measurement unit (Keithley 2450). The $R_s$ was measured by d.c. four probe method in the van der Pauw electrode configuration. The $S$ was measured by giving a temperature difference ($\Delta T$) of 5 K in the film using two Peltier devices, where the actual temperatures of both sides of SrCoO$_x$ layer were monitored by two tiny thermocouples. The thermo-electromotive force ($\Delta V$) and $\Delta T$ were simultaneously measured, and the $S$ was obtained from the slope of the $\Delta V$–$\Delta T$ plot.

## III. Results & Discussions

### A. SrCoO$_x$ TFT with leakage-free water

We first fabricated the SrCoO$_x$ TFT with leakage-free water. The 200-nm-thick CAN film was deposited at RT under $P_{O2}$ of 5 Pa by PLD, because the film density of the CAN film can be controlled by the $P_{O2}$ during the deposition and the film porosity ($\Phi$) at $P_{O2}$ = 5 Pa was 32 %.[25] The $\sigma$ of the CAN film was measured to be 2.0 × 10$^{-8}$ S cm$^{-1}$ at RT. Considering that the water is fully occupied in the nanopore region (32 % of the total volume of CAN film), the $\sigma$ of water in the CAN film is estimated to be 6.3 × $10^{-8}$ S cm$^{-1}$, which is consistent with 5.6 × $10^{-8}$ S cm$^{-1}$ of the ultrapure water.[26] **Figure 2(a)** shows the high-resolution TEM image of the cross-sectional SrCoO$_x$ TFT, where the electron incident direction was SrTiO$_3$ [110]. The multilayered structure composed of SrCoO$_{2.5}$ (30 nm) / CAN (160 nm) / Ti (20 nm) was observed and the numerous light spots with diameters of 10–20 nm are seen in the CAN film, indicating the presence of high-density nanopores.

The electrochemical switching of the SrCoO$_x$ TFT with leakage-free water was examined by measuring $R_s$ and $S$ at RT. **Figure 2(b)** shows the variation of $R_s$ with respect to the retention time ($t$) of various $V_g$, where the $R_s$ was measured immediately after the bias application for 10 minutes at each $V_g$. First, negative $V_g$ up to –20 V was applied for the oxidation of SrCoO$_{2.5}$ layer (left panel), and then positive $V_g$ up to +15 V was applied for the reduction (right panel). The $R_s$ decreased from 4.5 MΩ sq.$^{-1}$ at the initial state down to 0.5 kΩ sq.$^{-1}$ at $V_g$ = –20 V, due to the oxidation of SrCoO$_{2.5}$ film. Subsequently, by applying positive $V_g$ up to +15 V, the $R_s$ recovered to insulating state (46 MΩ sq.$^{-1}$) due to the reduction. **Figure 2(c)** summarizes the $S$ with respect to the sheet conductivity (1/$R_s$). The $\Delta V$–$\Delta T$ plots at RT are shown in the inset of **Fig. 2(c)**, which ensures a linear relationship between $\Delta V$ and $\Delta T$. The measured $S$ are always positive, indicating that the SrCoO$_x$ layer is a p-type conductor. The $S$ decreased form +190 μV K$^{-1}$ to +10 μV K$^{-1}$, which is almost the same with that of metallic SrCoO$_3$ phase.[14] Since the $S$ basically reflects the energy differential of the density of states (DOS) around the Fermi level ($E_F$), $\left[\frac{\partial \text{DOS}(E)}{\partial E}\right]_{E=E_F}$,[27,28] the decrease of $S$ with increase of 1/$R_s$ reflect the electronic-structure change from insulating to metallic state of SrCoO$_x$ layer by the oxidation. However, these results imply that the electrochemical switching of SrCoO$_x$ TFT with leakage-free water requires the application of high $V_g$ up to 20 V for a very long retention-time ~40 minutes, which should be necessary to gain the enough coulomb density for the electrochemical reaction of SrCoO$_x$ layer, primarily due to the low $\sigma$ of the leakage-free water.[29]

**B. Conductivity control of leakage-free electrolyte**

We then examined the conductivity control of leakage-free electrolyte by changing $x$ of a-Na$_x$TaO$_3$ films. The a-Na$_x$TaO$_3$ films with nominal $x$ = 0.01, 0.1, 1.0 were deposited at RT by PLD. **Figure 3(a)** summarizes the $\Phi$ of a-NaTaO$_3$ films with respect to $P_{O2}$ during the deposition. When the $P_{O2}$ exceeds 4 Pa, the plume becomes large, as shown in the inset. With increase of $P_{O2}$ > 5 Pa, the film density steeply starts to decrease, leading to the formation of porous a-NaTaO$_3$ film (**Supplementary Fig. 1** and **Fig. 2**) and the $\Phi$ reached to 60 % at $P_{O2}$ = 11 Pa. **Figure 3(b)** shows the bright field (BF) STEM image of the a-NaTaO$_3$ films fabricated at $P_{O2}$ of 4 Pa and 11 Pa. Although the a-NaTaO$_3$ film fabricated at low $P_{O2}$ of 4 Pa is fully dense, that grown under high $P_{O2}$ of 11 Pa has a nanopillar array structure. **Figures 4(a-c)** summarize the impedance plots at RT in air for the a-Na$_x$TaO$_3$ films ($x$ = 0.01–1.0) with nanopillar array structure, fabricated at $P_{O2}$ = 11 Pa. The semicircle was observed at high frequency region ≥ 10 kHz. The estimated $\sigma$ of a-Na$_x$TaO$_3$ films was largely controlled from $5.2 \times 10^{-8}$ S cm$^{-1}$ ($x$ = 0.01) to $1.2 \times 10^{-7}$ S cm$^{-1}$ ($x$ = 0.1) to $2.8 \times 10^{-6}$ S cm$^{-1}$ ($x$ = 1), compared to $2.0 \times 10^{-8}$ S cm$^{-1}$ of the CAN film, indicating that the change of $x$ in a-Na$_x$TaO$_3$ film realized the control of $\sigma$ of the leakage-free electrolyte.

**C. SrCoO$_x$-TFT using leakage-free electrolyte with different conductivity**

In order to compare the electrochemical switching properties of SrCoO$_x$ TFT using the gate insulator of a-$Na_xTaO_3$ films with different conductivity, we investigated the relationship between the flowing $I_g$ and $R_s$ under the $V_g$ applications. **Figure 5(a)** shows the variation of $I_g$ and $R_s$ with respect to the retention time ($t$) of $V_g = -3$ V. For all the SrCoO$_x$ TFTs with a-$Na_xTaO_3$ ($x = 0.01$–$1.0$) gate insulators, $-I_g$ increased with $t$ and the $R_s$ simultaneously decreased by applying the $V_g$, indicating the electrochemical oxidation of the SrCoO$_x$ layer. Meanwhile, the $t$ required for the metallization of 30-nm-thick SrCoO$_x$ layer becomes two orders of magnitude shorter with increase of the $\sigma$ of a-$Na_xTaO_3$ gate insulator. The time scale of the electrochemical switching can be considered to depend on ion migration in the electrolyte, electric double layer formation, electrochemical surface reaction, and $O^{2-}$ ion diffusion into SrCoO$_x$ layer under the $V_g$ application. Considering that the $\sigma$ of a-$Na_xTaO_3$ leakage-free electrolyte largely determine the time scale of the electrochemical switching, the switching strongly depends on the density of $OH^-$ ions accumulated at the SrCoO$_x$ surface. These results suggest that the ion migration in the leakage-free electrolyte is the rate-determining step for the device operation, compared to the other electrochemical process. Therefore, the high $\sigma$ of the leakage-free electrolyte is the key factor for the electrochemical switching of SrCoO$_x$ TFT.

It should be noted that the $R_s$ does not start to decrease until the $I_g$ increase up to a certain value. Since the $V_g$ is applied between the gate-source electrodes, the SrCoO$_x$ layer first becomes metallic around the source electrode, and then, the metallic region gradually expands toward drain electrode. Finally, the metallic parallel-plate structure is formed, and the $I_g$ effectively flows at the entire region, which decreases the $R_s$ of SrCoO$_x$ layer.

**Figure 5(b)** compares the variation of $R_s$ with electron density ($Q$), as calculated from the integral value of the $I_g$–$t$ plots in **Fig. 5(a)**. The $R_s$ decreased along the universal line and the oxidation processes were completed at ~$1.0 \times 10^{17}$ cm$^{-2}$, which corresponds with the ideal $Q$ of $1.1 \times 10^{17}$ cm$^{-2}$ required for the electrochemical oxidation form SrCoO$_{2.5}$ to SrCoO$_3$ (SrCoO$_{2.5}$ + $2x$OH$^-$ ⇆ SrCoO$_{2.5+x}$ + $x$H$_2$O + $2x$e$^-$ ($0 \leq x \leq 0.5$)). The universal change in $R_s$, regardless of the $\sigma$ of a-Na$_x$TaO$_3$ gate insulator, indicates that all the provided electrons were used for the electrochemical redox reaction of the SrCoO$_x$ layer, obeying Faraday's laws of electrolysis, and that the device operation can be controlled by current density.

## IV. Conclusions

In summary, the electrochemical switching of SrCoO$_x$ TFTs was examined by using the leakage-free electrolytes with different $\sigma$ as the gate insulator. We first examined the SrCoO$_x$ TFT with water-infiltrated CAN gate insulator, but the switching required the application of high $V_g$ up to 20 V for a very long $t$ = ~40 minutes, primarily due to the low $\sigma$ of the leakage-free water. We then fabricated the a-Na$_x$TaO$_3$ film with different $x$ = 0.01–1.0 and demonstrated the control of $\sigma$ from $8.0 \times 10^{-8}$ S cm$^{-1}$ ($x$ = 0.01) to $2.5 \times 10^{-6}$ S cm$^{-1}$ ($x$ = 1) at RT, compared to the $2.0 \times 10^{-8}$ S cm$^{-1}$ of the CAN film. As the result, the $t$ at small $V_g$ = –3 V, required for the metallization of SrCoO$_x$ layer, becomes two orders of magnitude shorter with increase of the $\sigma$ of the a-Na$_x$TaO$_3$ leakage-free electrolytes. These results indicate that the ion migration in the leakage-free electrolyte is the rate-determining step for the device operation, compared to the other electrochemical process, and the high $\sigma$ of the leakage-free electrolyte is the key factor, leading to the development of the non-volatile electro-magnetic phase switching device of SrCoO$_x$ TFT.

**Acknowledgments**

We thank N. Hirai for the TEM/STEM analyses. The TEM/STEM analyses, conducted at Hokkaido Univ., were supported by Nanotechnology Platform Program from MEXT. TK was supported by PRESTO, JST (JPMJPR16R1), Grant-in-Aid for Young Scientists A (15H05543) and Grant-in-Aid for Challenging Exploratory Research (16K14377) from JSPS. HO was supported by Grant-in-Aid for Scientific Research A (25246023) and Grant-in-Aid for Scientific Research on Innovative Areas (25106007) from JSPS.

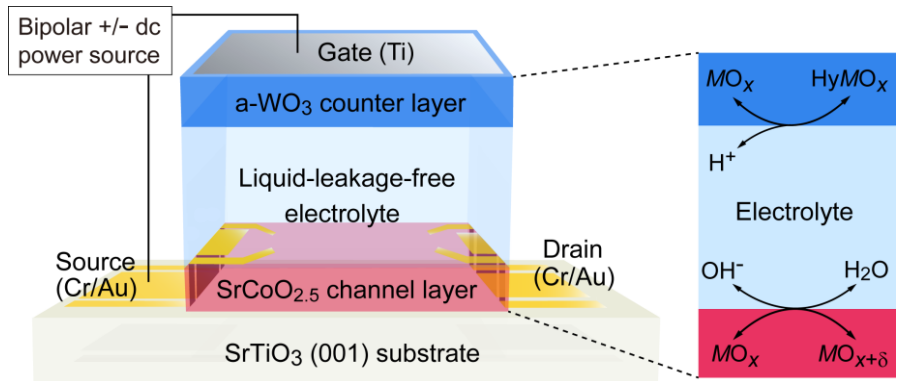

**FIG. 1.** A schematic of three-terminal TFT structure consisting of a $SrCoO_{2.5}$ channel layer, a liquid-leakage-free electrolyte as a gate insulator, an a-$WO_3$ counter layer, and a Ti gate electrode on a $SrTiO_3$ (001) substrate. Cr/Au bilayer films were used for source-drain electrodes. Right figure illustrates the electrochemical reaction in the device during the gate voltage application.

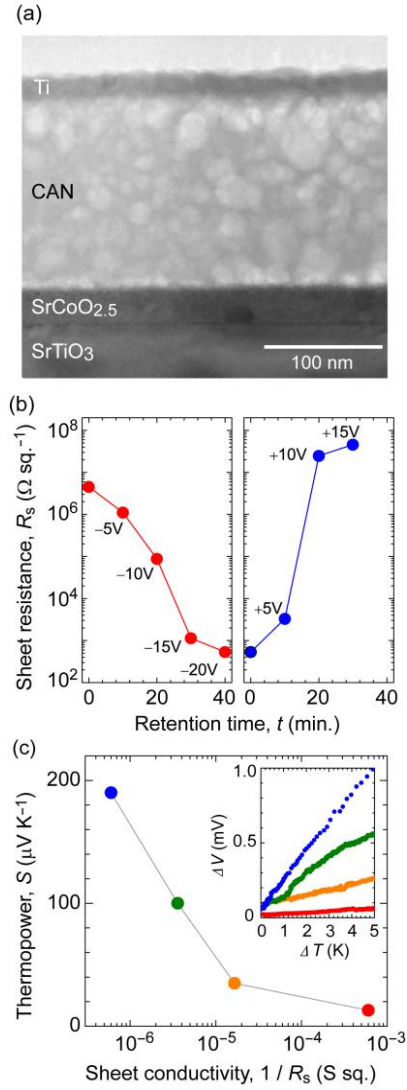

**FIG. 2.** (a) Cross-sectional TEM image of the SrCoO$_x$ TFT with leakage-free water. Trilayer structure composed of SrCoO$_{2.5}$ (30 nm), CAN (160 nm), and Ti (20 nm) layers is clearly seen. Many light spots in the CAN layer indicate nanopores, which is fully occupied with water. (b) $R_s$ vs. retention time ($t$) of applied $V_g$ for SrCoO$_x$ layer. The $R_s$ was measured after the $V_g$ application, where negative $V_g$ up to –20 V was applied for oxidation (left panel) and positive $V_g$ up to +15 V was applied for reduction (right panel) of SrCoO$_x$ layer. The $V_g$ application time was 10 minutes at each step. (c) $S$ vs. 1/$R_s$ at RT. $\Delta V$–$\Delta T$ plots are shown in the inset.

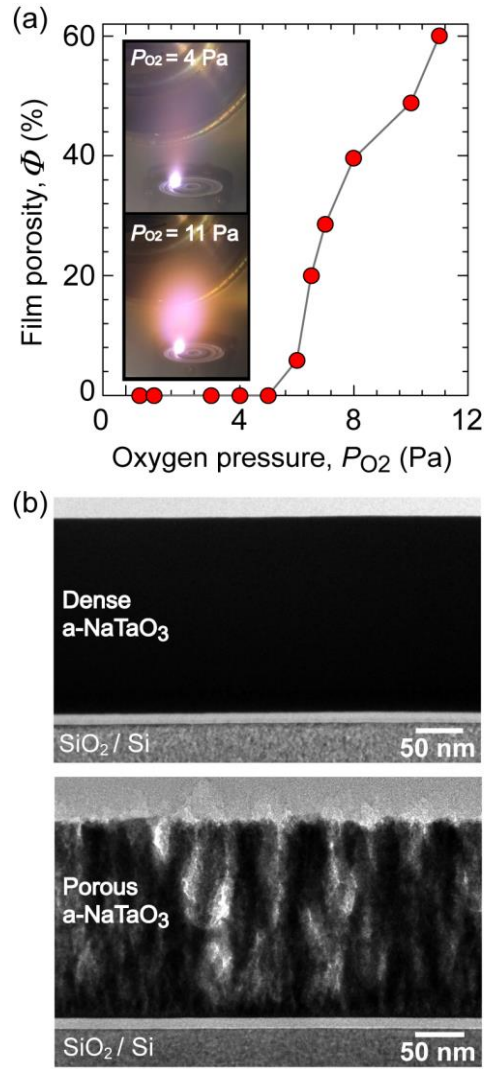

**FIG. 3.** Microstructure analyses of a-NaTaO$_3$ film. (a) Film porosity ($\Phi$) as a function of oxygen pressure ($P_{O2}$) during the deposition at RT. The insets show the pictures of ablation plume of NaTaO$_3$ under $P_{O2}$ = 4 Pa and 11 Pa. (b) BF-STEM images of dense a-NaTaO$_3$ film (upper panel) and porous a-NaTaO$_3$ film (lower panel), deposited on SiO$_2$/Si substrates at $P_{O2}$ = 4 Pa and 11 Pa, respectively.

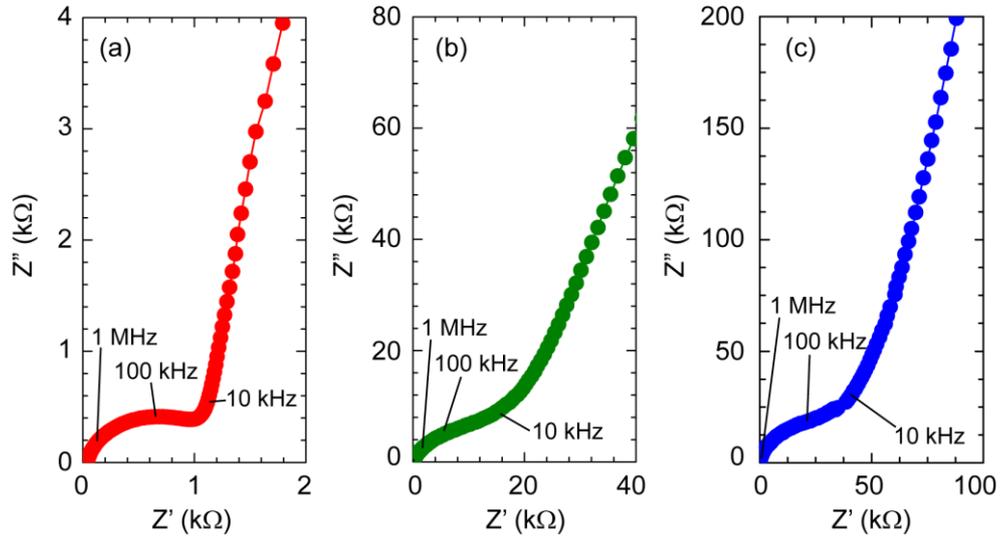

**FIG. 4.** Cole-Cole plots of 300-nm-thick a-Na$_x$TaO$_3$ films (950 μm × 800 μm) with $x$ = 1.0 (a), 0.1 (b), and 0.01 (c), measured at RT in air.

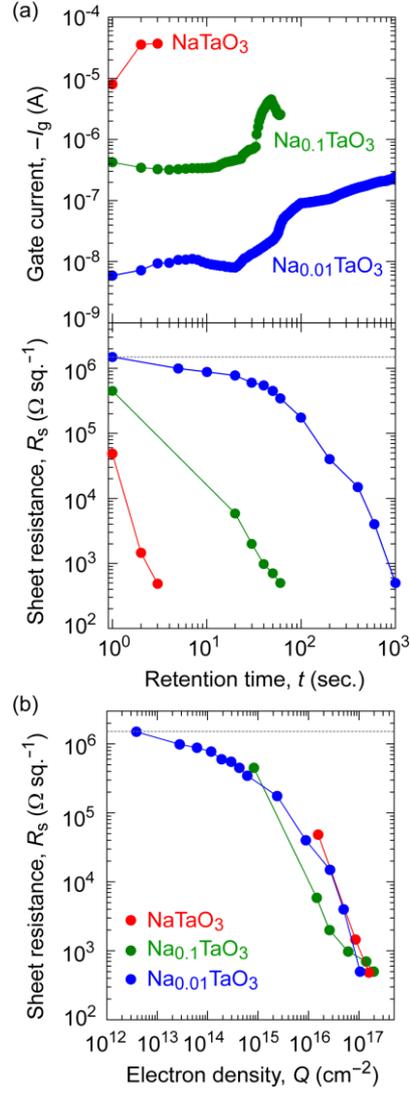

**FIG. 5.** (a) Retention-time ($t$) dependence of $I_g$ (upper panel) and $R_s$ (lower panel) for the SrCoO$_x$ TFT with a-Na$_x$TaO$_3$ ($x$ = 0.01, 0.1, 1.0) leakage-free electrolyte under the application of $V_g$ = −3 V. The $I_g$ and $R_s$ were respectively measured during and after the $V_g$ application. The dotted lines indicate the $R_s$ of the virgin SrCoO$_{2.5}$ layer. (**b**) Electron-density ($Q$) dependence of $R_s$ under the $V_g$ application. The $Q$ was calculated as the integrated value of the $I_g$–$t$ plots in (a).

# Leakage-free electrolytes with different conductivity for non-volatile memory device utilizing insulator/metal ferromagnet transition of SrCoO$_x$

Takayoshi Katase[1,2,3,a)], Yuki Suzuki[4], and Hiromichi Ohta[1,a)]

[1] *Research Institute for Electronic Science, Hokkaido University, N20W10, Kita, Sapporo, 001-0020, Japan*

[2] *Laboratory for Materials and Structures, Institute of Innovative Research, Tokyo Institute of Technology, 4259 Nagatsuta, Midori, Yokohama, 226−8503, Japan*

[3] *PRESTO, Japan Science and Technology Agency, 7 Gobancho, Chiyoda, Tokyo, 102-0076, Japan*

[4] *School of Information Science and Technology, Hokkaido University, N14W19, Kita, Sapporo, 060−0814, Japan*

a) Correspondence and requests for materials should be addressed to: T. Katase (katase@mces.titech.ac.jp) and H. Ohta (hiromichi.ohta@es.hokudai.ac.jp)

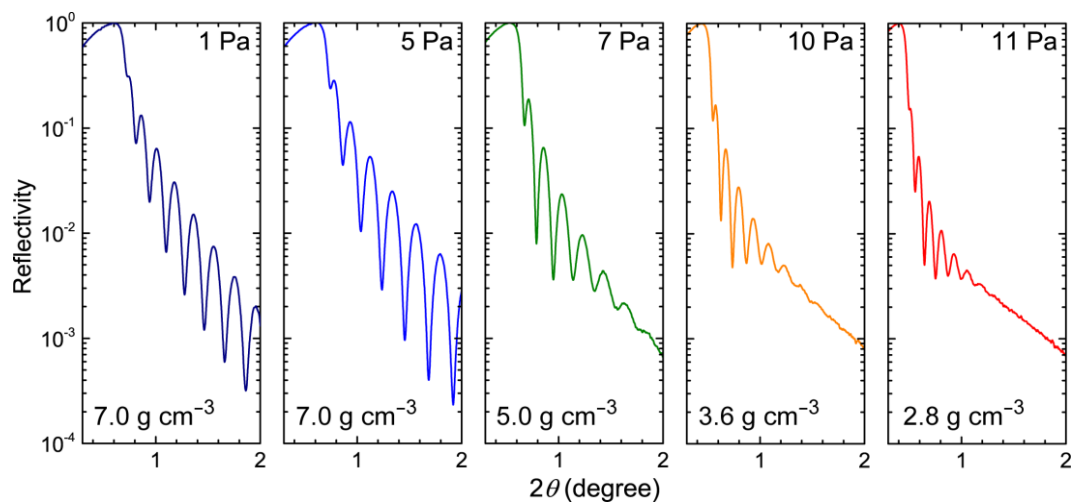

**Supplementary FIG. S1.** Grazing-incidence X-ray reflectivity of the a-NaTaO$_3$ films deposited under various $P_{O2}$ = (a) 1 Pa, (b) 5 Pa, (c) 7 Pa, (d) 10 Pa, and (e) 11 Pa. The film density largely depends on the $P_{O2}$ [(a) 7.0 g/cm$^3$, (b) 7.0 g/cm$^3$, (c) 5.0 g/cm$^3$, (d) 3.6 g/cm$^3$, (e) 2.8 g/cm$^3$]. The film density of 7.0 g/cm$^3$ at low $P_{O2}$ = 1–5 Pa agrees well with that (7.0 g/cm$^3$) of NaTaO$_3$ bulk, but significantly decreased to less than half with increase of $P_{O2}$ up to 11 Pa.

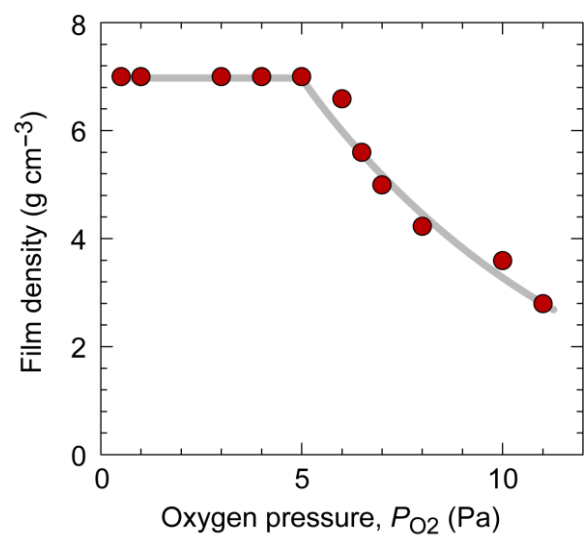

**Supplementary FIG. S2.** Film density of a-NaTaO$_3$ films as a function of oxygen pressure ($P_{O2}$) during the deposition at RT.